\documentclass[12pt,aps,floatfix]{revtex4}
\makeatletter
\gdef\@ptsize{0}
\gdef\@ptsize{1}
\gdef\@ptsize{2}
\let\@currsize\normalsize
\makeatother
\usepackage{setspace}
\usepackage{amsmath}    
\usepackage{graphicx}   
\usepackage{verbatim}   
\usepackage{color}      
\usepackage{subfig}  
\usepackage{hyperref}   
\usepackage{setspace}
\usepackage{pdflscape}
\usepackage{multirow}
\usepackage{verbatim}
\usepackage{changepage}

\usepackage{algpseudocode}
\usepackage{algorithm}
\usepackage{times}
\usepackage{float}
\floatstyle{plaintop}
\restylefloat{table}
\usepackage{cases}
\usepackage{hyperref}
\usepackage{wasysym}
\usepackage{braket}
\usepackage[utf8]{inputenc}
\usepackage[T1]{fontenc}
\usepackage{todo}

\newcommand{\bvec}[1]{{{\textbf{{#1}}}}}
\begin{document}

\title[(Non-)Magnetization of 433 Eros]{The (Non-)Magnetization of 433 Eros: Possible Mechanisms for the Lack of Magnetism as Measured by NEAR}
\author{Niraj K. Inamdar}
\date{6 December 2013}
\begin{abstract}
The Near Earth Asteroid Rendezvous-Shoemaker (``NEAR'') spacecraft orbited and ultimately landed on the near-Earth asteroid 433 Eros. One of the primary science objectives of NEAR was the MAG experiment, which measured the magnetic field in the vicinity of Eros during orbit and after landing. Eros is therefore at the present the best characterized asteroid using \textit{in situ} measurement of magnetism. MAG results suggested that Eros was very unmagnetized---with an upper bound on the natural remanent magnetism (NRM) placed at $1.9\times 10^{-6}\mathrm{A\cdot m^2 \cdot kg^{-1}}$---especially when compared to meteorite samples of analogous composition. Since meteorites and asteroids are typically believed to represent the remnants of disrupted parent bodies, the ramifications of the low level of magnetization of Eros are considerable, since it could imply disparate origins for objects of similar composition. In this paper, we explore whether there are any systematic effects related to the actual process of measurement and derivation of the Erotian NRM, and whether such effects played a role in the low levels of NRM derived for Eros. By simulating the orbit of NEAR around Eros and using the field strength values measured by NEAR, we find that we are able to place a higher bound on the NRM of Eros by a factor of at least an order of magnitude higher than that originally suggested by Acu\~{n}a, \textit{et al.} (2002). We find that if we suppose Eros to be made up of constituents that have roughly uniform magnetization directionally, that it is possible to infer an L or LL chondrite-type composition for Eros within the bounds of values for remanent magnetism reported in the meteorite record. The results provide a more rigorous confirmation of the suggestion by Wasilewski, \textit{et al.} (2002) that Eros cannot be ruled out as an L- or LL-type analogue.
\end{abstract}
\maketitle

\section{Introduction} 
Between February 2000 and February 2001, the Near Earth Asteroid Rendezvous-Shoemaker (hereafter ``NEAR'') probe orbited near-Earth asteroid 433 Eros before ultimately landing on it. The primary science objectives of NEAR were to characterize the composition and bulk physical properties of Eros (Table \ref{tab:PhysPropsEros}). The scientific payload of NEAR comprised the Multispectral Imager (MSI), Near-Infrared spectrograph (NIS), X-ray Spectrometer (XRS), Gamma Ray Spectrometer (GRS), Magnetometer (MAG), NEAR Laser Rangefinder (NLR), and Radio Science (RS). In concert, these instruments characterize the physical properties of the asteroid, potentially shedding light on its evolutionary history and its possible link to meteorite specimens on Earth. 

The conventional wisdom is that many asteroids and meteorites share a common origin as remnants of disrupted parents bodies that have since been fragmented as a result of a complex evolutionary history that may include collisions and impacts. This is supported by spectral and isotopic analyses, in which specimens are compared to one another and grouped based on observed similarities, such as elemental abundance and isotope ratios. Likewise, the existence of asteroid families (asteroids with similar orbital elements) in which constituent members share similar spectral signatures lends itself to this picture of parent body disruption. We might then expect meteorites and asteroids to share other physical similarities, such as bulk magnetization. 

Magnetization is recognized as a diagnostic for the evolutionary history of a terrestrial body \cite{tauxe2010essentials, kaula1968introduction, weiss2013differentiated}. In particular, remanent magnetism serves as an indicator of the magnetic fields to which a particular physical object was once exposed. Within the context of dynamo theory, this can allow us to probe the temporal evolution of a dynamo on a hypothetical parent body. Furthermore, if a hypothetical parent body contained a dynamo, the knowledge of its existence would place direct constraints on the physical properties of the parent body and its possible thermal evolution history \cite{fu2012ancient}. The widespread phenomenon of magnetism in chondrites---believed to have never experienced melting before accretion onto their parent bodies---may in fact have an origin in internally differentiated parent bodies that were overlain with chondritic crusts. \cite{anders1961theories,carporzen2011magnetic,weiss2013differentiated}. 

\begin{table}[t]
	\caption{Key Physical Properties of 433 Eros\label{tab:PhysPropsEros}}
		\begin{tabular}{|c|c|}\hline
		Property                     & Value  \\ \hline \hline
				Size         		        		& $34.4 \times 11.2 \times 11.2~\mathrm{km}$ \cite{chamberlin2010jpl,zuber2000shape}\\ \hline
		    Mass          		      		& $6.69 \times 10^{15}~\mathrm{kg}$ \cite{chamberlin2010jpl}         \\ \hline
			  Mean density   		      		& $2.67\pm 0.03~\mathrm{g \cdot cm^{-3}}$ \cite{chamberlin2010jpl}    \\ \hline
				Asteroid                		&	\multirow{2}{*}{S-type (SMASS) \cite{chamberlin2010jpl}}	\\ 
				spectral  class             &																						 \\ \hline						
				Analogue                		&	$\sim$H/ordinary chondrite                	 \\ 
				meteorite class             &	(depleted in S) \cite{lim2009elemental,nittler2001x} 	 \\ \hline									
		\end{tabular}
\end{table}

\begin{figure}[ht]
\begin{center}
\subfloat[Near infrared and visible spectra of Eros from 3 different observations. Data credit: SMASS \cite{bus2002phase,burbine2002small,SpeX}.
]{\includegraphics[width=0.475\textwidth]{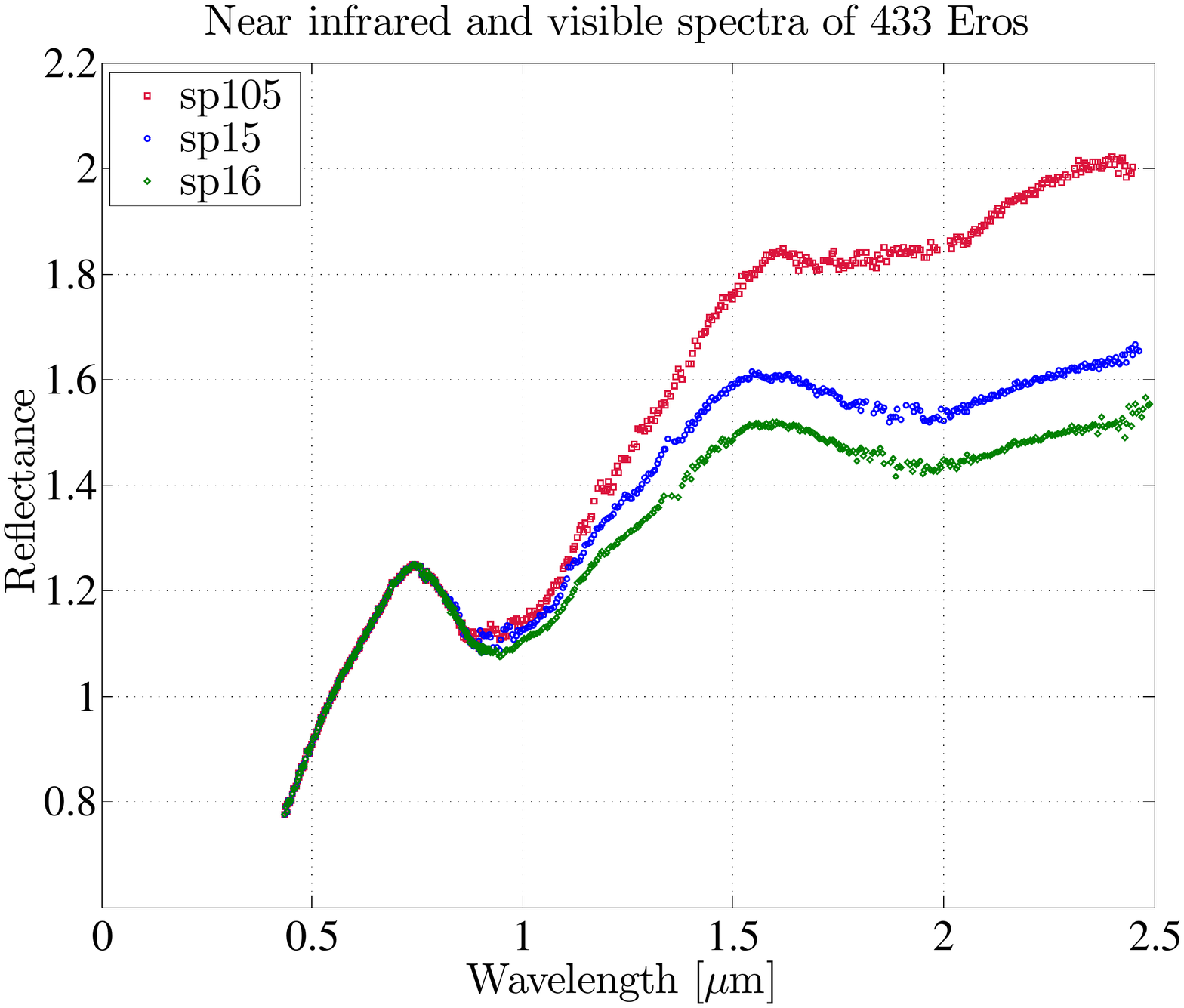} \label{fig:ErosSpectra}}
\subfloat[Comparison of visible spectra of Gaspra and Eros. Data credit: SMASS \cite{bus2002phase,burbine2002small,SpeX}. 
]{\includegraphics[width=0.475\textwidth]{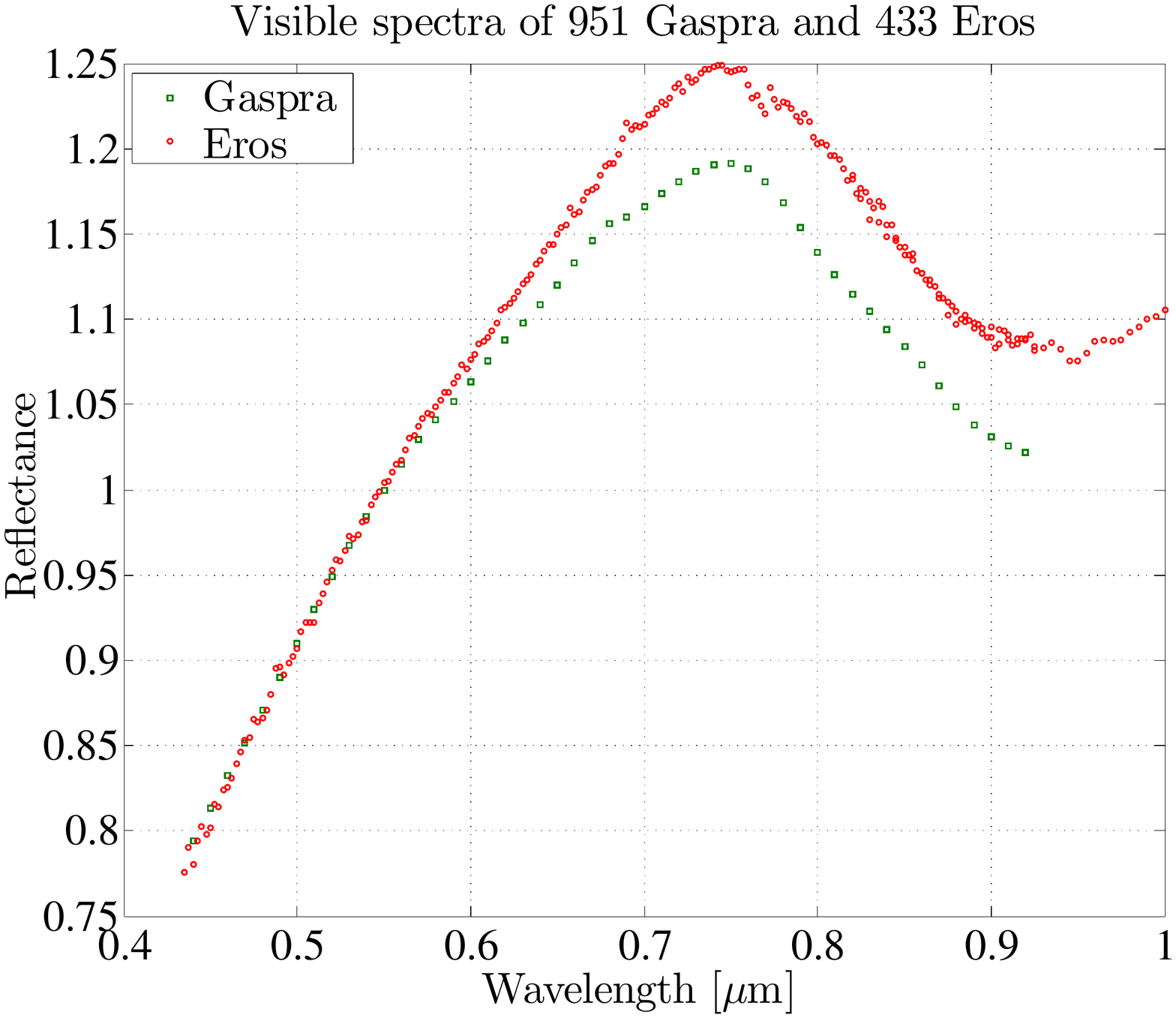} \label{fig:ErosGaspComp}}
\end{center}
\caption{Visible and near infrared spectra of Eros and Gaspra.}
\label{fig:VandNIRSpectra}
\end{figure}

\subsection{Possible Magnetism on Eros} 
In 1991, the \textit{Galileo} spacecraft made the first ever close approach to an asteroid when it passed within about 1,600 km of 951 Gaspra, another S-type asteroid \cite{kivelson1997europa}. From some 230 Gaspra radii away, \textit{Galileo} detected what was believed to be evidence of a magnetic field; its presence was inferred from a ``draping'' of the solar wind by what could be a magnetosphere. Corresponding to a field strength of $1.4 \times 10^{-4}~\mathrm{T}$, an upper bound on the magnetic moment for Gaspra was placed at $2\times 10^{14}~\mathrm{A\cdot m^2}$ (for reference, the value for Earth $M_{\earth} \approx 8 \times 10^{22}~\mathrm{A\cdot m^2}$) \cite{kivelson1997europa}. Kivelson, \textit{et al.} thus placed an upper bound on the NRM of Gaspra at $3\times 10^{-2}~\mathrm{A\cdot m^2 \cdot kg^{-1}}$. 

The results of Kivelson, \textit{et al.} have since been strongly questioned \cite{blanco2003hybrid}. Simulations of the interaction of a magnetic dipole with the solar wind by Blanco-Cano, \textit{et al.} have concluded that a field strength of $3.3 \times 10^{-6}~\mathrm{T}$ was in fact responsible for the signatures seen in \textit{Galileo}'s measurements. A straightforward calculation shows that the corresponding NRM for Gaspra would then be $7.3\times 10^{-4}~\mathrm{A\cdot m^2 \cdot kg^{-1}}$, some 40 times smaller than the value given by Kivelson, \textit{et al.}  
Gaspra's spectral features (Fig. \ref{fig:ErosGaspComp}) suggest that, while it is an S-type asteroid, it is somewhat richer in olivine than other S-types \cite{chapman1996s}. While it was originally believed that Gaspra was metal-enriched \cite{kivelson1993magnetic}, this interpretation was later discounted \cite{chapman1996s} and no metal-rich material was found from the \textit{Galileo} images \cite{kivelson1993magnetic}. Based on the Small Main-Belt Asteroid Spectroscopic Survey (SMASS) classification scheme \cite{demeo2009extension}, Eros is also an S-type asteroid, its spectral features (see Fig. \ref{fig:ErosSpectra}) typically associated with iron- and magnesium-bearing oxides. 

Due to the spectral similarity between Eros and Gaspra, one of the primary goals of NEAR was therefore to determine whether Eros was similarly magnetized \cite{acuna2002near}; an additional incentive was provided when Richter, \textit{et al.} also reported relatively high intrinsic levels of magnetism on the asteroid 9969 Braille as measured by the Deep Space 1 probe \cite{richter2001first}. The value determined by Richter, \textit{et al.} for the NRM of Braille was $2.8\times 10^{-2}~\mathrm{A\cdot m^2 \cdot kg^{-1}}$. We note, however, the relatively poor quality of the fit achieved in determining this magnetic moment---see Fig. 3, top panel, of Richter, \textit{et al.} \cite{richter2001first}---which again raises some questions about this relatively high level of remanent magnetism. The interest in Braille stems from the fact that it is a Q-type asteroid, a classification which has been associated with ordinary chondrites \cite{sasaki2001production} and S-type asteroids \cite{mcsween1991mineralogy}. 

\section{Review of NEAR Results}	
\subsection{NEAR Spectral Results \label{sec:NearSpectral}}
NEAR's X-ray spectral analysis has suggested that Eros has characteristics similar to an ordinary H chondrite \cite{weisberg2006systematics}, albeit with depleted S/Si ratios, which Lim and Nittler \cite{lim2009elemental} and Nittler, \textit{et al.} \cite{nittler2001x} suggest may be due to space weathering-induced sputtering of the Erotian regolith or impact volatilization. Nittler, \textit{et al.} and Trombka, \textit{et al.} suggest that the observed depletion on the Erotian regolith of sulfur compared to other ordinary chondrites may be due to the loss of an FeS-rich partial melt; differentiation globally is ruled out \cite{nittler2001x,trombka2000elemental}.

\subsection{MAG Results}

The NEAR magnetic field investigation MAG performed extensive magnetic field measurements of Eros over a broad range of distances. This range encompassed the cruise and approach phases of the mission ($100,000~\mathrm{km}$ to $400~\mathrm{km}$), the primary observation phase after orbital insertion (February 2000 to February 2001), and the actual landing of the spacecraft on the surface. 

Throughout the approach and orbital phase, no field indistinguishable from the interplanetary magnetic field (IMF) was detected, with a typical field strength of $\sim 1~\mathrm{nT}$ \cite{acuna2002near}. Measurements on the surface, which occurred over a period of about two weeks, give an upper bound on the magnetic flux density at about $5~\mathrm{nT}$ \cite{acuna2002near}. From this number, Acu\~{n}a, \textit{et al.} derived an upper bound on the bulk NRM of Eros of $1.9\times 10^{-6}~\mathrm{A \cdot m^2 \cdot kg^{-1}}$, some four orders of magnitude less than what was at the time calculated to be the NRM value for Gaspra.

\section{Intrinsic Reasons for the Low Magnetization of Eros}

\begin{figure}[ht]
\begin{center}
\includegraphics[width= 0.75\textwidth]{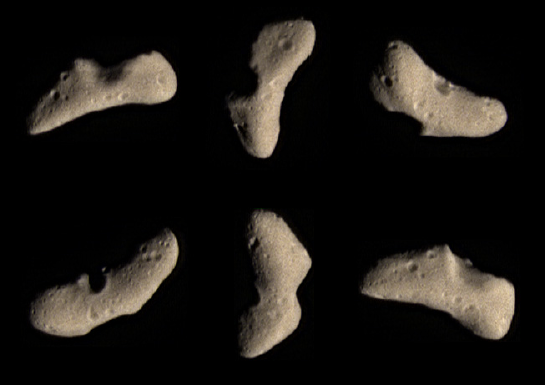}
\caption{Some views of Eros from the NEAR-Shoemaker mission. Psyche, Eros's second-largest crater, is see most clearly on the view on the lower left, while Himeros, its largest, is the large depression seen most clearly in the upper three views (from left to right, on the top, left, and lower left of Eros). Figure credit: NASA\label{fig:Eros433}.}
\end{center}
\end{figure}		
\subsection{Collisional History}	
Impact-induced shock and melt demagnetization is a possible explanation for Eros's low magnetization. Many meteorites bear evidence of shock and melt, and if these processes occur in a non-magnetic environment, then there is a possibility for partial demagnetization \cite{sugiura1988magnetic}. Shock will typically result in demagnetization factors of order unity. For the demagnetization levels observed on Eros, we would then expect substantial evidence of such collisions, perhaps of a catastrophic nature. There are indeed several interesting geological features on Eros, including two large craters, Psyche and Himeros (Fig. \ref{fig:Eros433}), and what is believed to be a thrust fault (Hinks Dorsum) connecting the two \cite{watters2011thrust}. Furthermore, the bimodal shape of Eros led to the initial suggestion that it represented the fusion of two other bodies. However, gravity field measurements \cite{zuber2000shape} and porosity calculations \cite{wilkison2002estimate} suggest that Eros is fairly structural and have ruled out the fusion scenario. Rough calculations carried out by the author based on previously-reported scaling relationships \cite{mohit2004impact} suggest that the sizes of the Eros's most prominent impact features are likely not commensurate with global demagnetization due to impact. 

\subsection{Vector Subtraction of Randomized Constituents}
Another possible explanation for the lack of magnetism on Eros is that it is composed of randomly-oriented constituents, each of which may be magnetized on a small scale (say $\mathrm{\mu m}$-scale) but which, when averaged over the entire asteroid, lead to an effective non-magnetization due to the vector subtraction of the individual components. That this is the case is reflected in part in our simulations (see Table \ref{tab:MeasNRMLong}). Indeed, magnetism in meteorites is known to have some dependence on sample size, with larger samples showing systematically lower levels of magnetism due to such an effect \cite{wasilewski2002443}. In this scenario, individual components may have acquired their magnetism earlier in their histories when a different solar state meant higher levels of magnetism in the interplanetary field. These individual components would then clump together randomly. Of course, without having kimometer- or Erotian-scale samples, or high-fidelity measurements of magnetism on other asteroids, it is impossible to say precisely what the trend is. 

\section{Systematic Reasons for the Low Magnetization of Eros}
\subsection{Problems with Meteorite Magnetism Record}
\begin{figure}[ht]
\begin{center}
\subfloat[ A comparison of the magnetization values reported by various studies for chondrite meteorites as indicated at the bottom of the figure. The labels ``S'' and ``L'' refer to small ($< 1~\mathrm{g}$) and large ($23,700~\mathrm{g}$) fragments of the meteorite ALH76009, an L6 chondrite. The ``(AF)'' refer to these samples after they were demagnetized in a $5~\mathrm{mT}$ alternating field.
]{\includegraphics[width=0.475\textwidth]{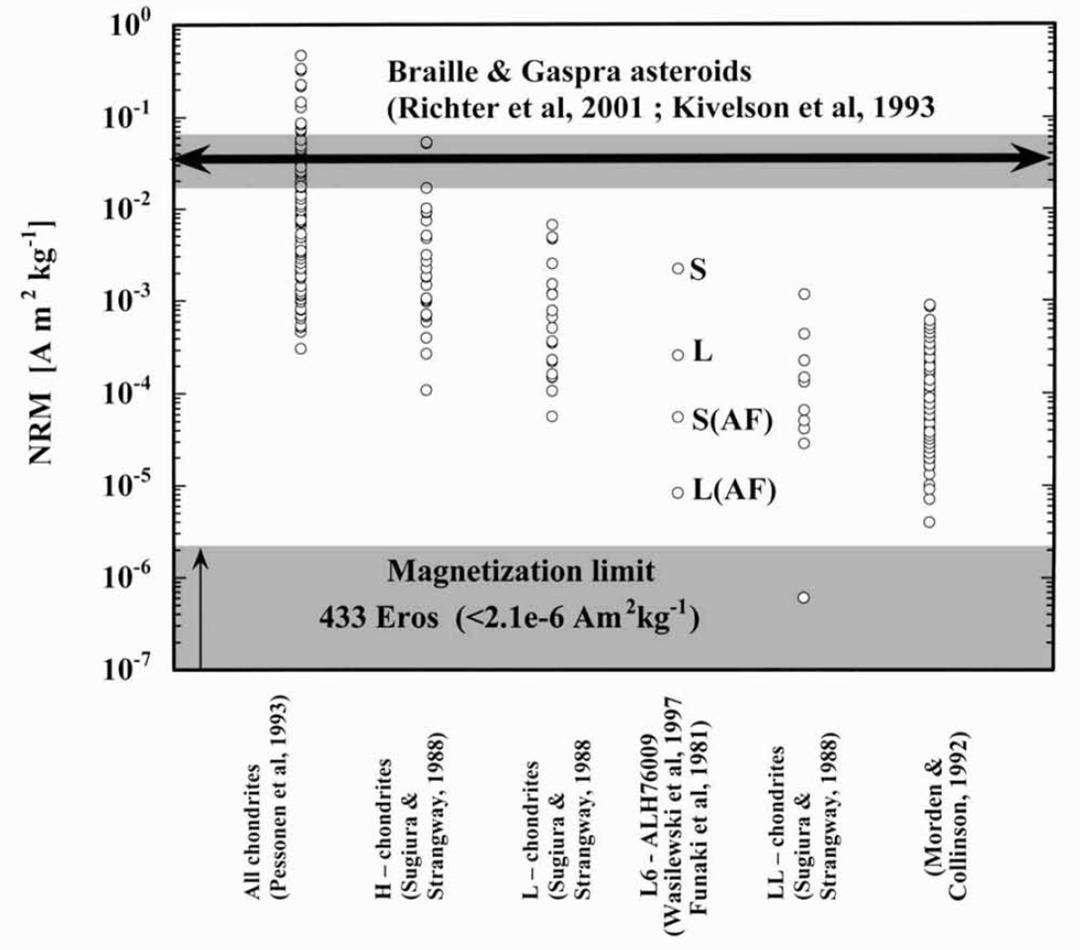} \label{fig:AlanHillsComp}}
\subfloat[Comparison of Eros magnetization limit with laboratory specimens when accounting for possible systematic errors in the meteorite magnetization record. The abscissa is the saturation remanence magnetization (SIRM), and the ordinate is the natural remanent magnetization (NRM) of various samples, with the Gaspra, Braille, and Eros limits indicated. The diagonal lines correspond to different values of the so-called ``REM'', or NRM:SIRM ratio.
]{\includegraphics[width=0.475\textwidth]{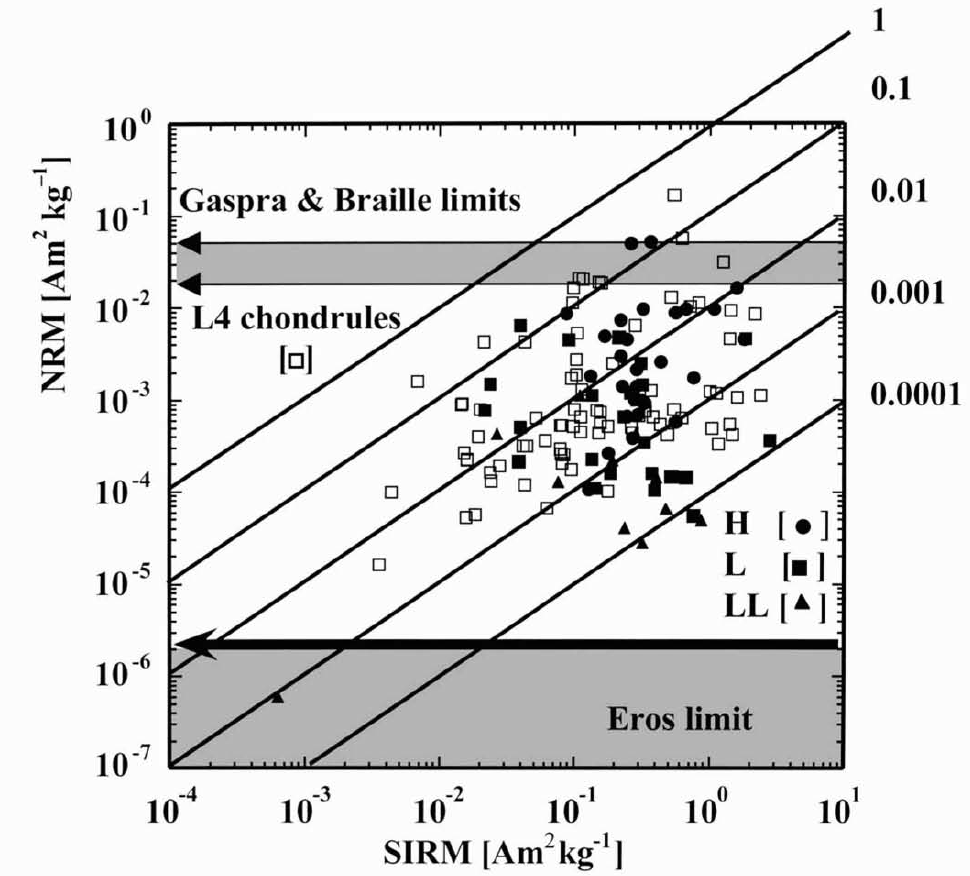} \label{fig:WasilewComp}}
\end{center}
\caption{Comparisons of the magnetism from various chondritic specimens. Figures from Wasilewski, \textit{et al.} \cite{wasilewski2002443}.}
\label{fig:MeteoriteSpecimenComps}
\end{figure}

As meteorites are the baseline against which asteroid magnetization measurements are compared, it is essential that the measurements made of meteorites are of the highest fidelity and represent the true remanant magnetism of the sample. However, it is often the case that meteorites are contaminated by terrestrial sources of magnetism such as hand magnets used to assess the possible extraterrestrial origin of the meteorite \cite{weiss2010paleomagnetic}. Thus, Wasilewski, \textit{et al.} suggested that the low level of magnetization in Eros should not necessarily be construed to be evidence of something remarkable \cite{wasilewski2002443}. In fact, according to the authors, the possible discrepancy in the expected magnetic field of Eros (based on spectral similarity) is in fact due to the meteoritic magnetization record against which Eros's magnetism is being compared. In Fig. \ref{fig:WasilewComp}, we see the meteorite record compared to the regions considered bounds for the magnetization of Gaspra, Braille, and Eros. Wasilewski \textit{et al.} have argued that the lowest L and LL chondrite specimens shown in the figure rules out precluding the interpretation of Eros as an LL chondrite analogue, particularly when it is clear that measurements of samples of similar or identical provenance have resulted in very different reports for NRM in the literature. 

\section{Numerical Investigation of Observational Effects of Magnetization Measurements}

MAG measured Eros's magnetic field throughout its primary science gathering phases, from orbit and down to its surface. Pending the landing of Rosetta on the comet 67P/Churyumov–Gerasimenko in mid 2014 \cite{glassmeier2007rosetta,auster2007romap}, NEAR's measurements are essentially unique magnetometer readings of a planetesimal. Hence, is worthwhile to understand whether there are observational reasons for NEAR's small field measurements. In this regard, we performed several numerical experiments to determine the field measurement dependence on factors such as orbit, and dipole orientation and strength. We first explore the phase space by considering varying levels of magnetization that span the values for ordinary chondrites as evidenced in the meteorite record, and assign those to the modeled constituent elements of Eros. We then focus on reconstructing a bulk NRM that is reflective of the field value limits given by NEAR. We then discuss the relevance of the results to the measurements made by NEAR and the possible implications of these results. 

Consider a magnetic dipole $d\bvec{m}$ at location $\bvec{r}'$. Then $d\bvec{m} = d\bvec{m}/dV'.dV'$, where $dV' = d^3\bvec{r}'$. The magnetic flux density $\bvec{B}$ due to all dipoles in the assembly as measured at location $\bvec{r}$ is the following integral over the volume $V'$: 
\begin{align}\label{eq:Hfield}
\bvec{B}(\bvec{r}) = \int d^3\bvec{r}' \frac{\mu_0}{4\pi}\left\{\frac{3\left(\bvec{r}-\bvec{r}'\right) \left[\left(d\bvec{m}/dV'\right) \cdot \left(\bvec{r}-\bvec{r}'\right)\right]}{\left|\bvec{r}-\bvec{r}' \right|^5} - \frac{d\bvec{m}/dV'}{\left|\bvec{r}-\bvec{r}' \right|^3}\right\}.
\end{align}
where $\mu_0$ is the permeability of free space. The field strength $\bvec{H} = \bvec{B}/\mu_0$. For us, $\bvec{r}$ always refers to the position of the NEAR spacecraft at a given time, and $V'$ refers to the volume of Eros. We therefore will investigate how the orientation and strength of these dipoles will drive measurements of the magnetic field in the vicinity of Eros as a function of NEAR's orbit, and how that may impact the determination of the bulk NRM.

\subsection{Orbit}
Due to the strong dependence of the measured field on the orbit, we first determine the approximate orbital radius $r_{\mathrm{det}}$ within which we would even expect to detect a magnetic signature. Denoting the mass of Eros by $M_{\textrm{Eros}}$, for a given bulk asteroid NRM and field detection value $|\bvec{B}|^*$, the orbital radius for a point-source Eros is approximately
\begin{align}
r_{\mathrm{det}} \approx \sqrt[3]{\frac{\mu_0}{2\pi}\frac{M_{\textrm{Eros}}\mathrm{NRM}}{|\bvec{B}|^*}}.
\end{align}

\begin{figure}[ht]
\begin{center}
\includegraphics[width=0.85\textwidth]{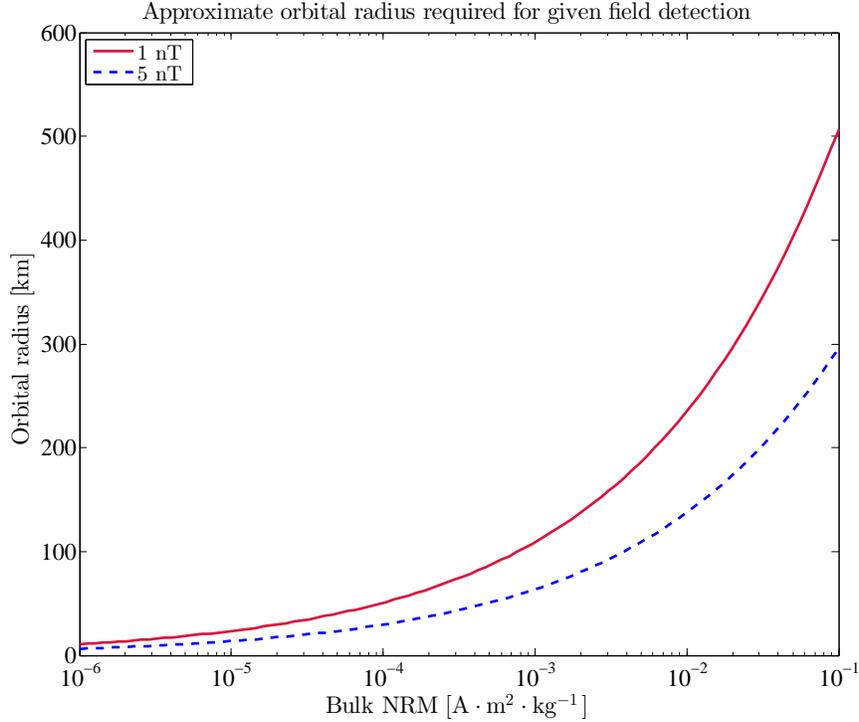}
\caption{Approximate orbital radius required for $1~\mathrm{nT}$ and $5~\mathrm{nT}$ field strength detection for various bulk NRM values.}
\label{fig:Orbit_reqt} 
\end{center}
\end{figure}	

In Fig. \ref{fig:Orbit_reqt}, we plot the orbital radii required for $1$ and $5~\mathrm{nT}$ flux densities (those reported as bounds for both orbit and surface measurements at Eros by Acu\~{n}a, \textit{et al.} \cite{acuna2002near}) for a range of NRM that span the ordinary chondrite values shown in Fig. \ref{fig:MeteoriteSpecimenComps}. For the NRM range within which the majority of ordinary chondrite specimens reside, the orbit should be $\lesssim 100~\mathrm{km}$.

Since the only period in which NEAR has a planar orbit at radii less than $100~\mathrm{km}$ (see the discussion in Sec. \ref{sec:OrbitNRM} for why we require a planar orbit) is from roughly 15 December 2000 until the end of the mission (12-13 February 2001), this is the time period over which we perform our numerical experiments. Orbit data for our simulations are taken from the NEAR SPICE data \cite{SPICE_PDS}. In our calculations, we always consider calculations in an Eros-fixed coordinate system.

\subsection{Eros Shape Model Geometry}
\begin{figure}[ht!]
\begin{center}
\includegraphics[width=0.6\textwidth]{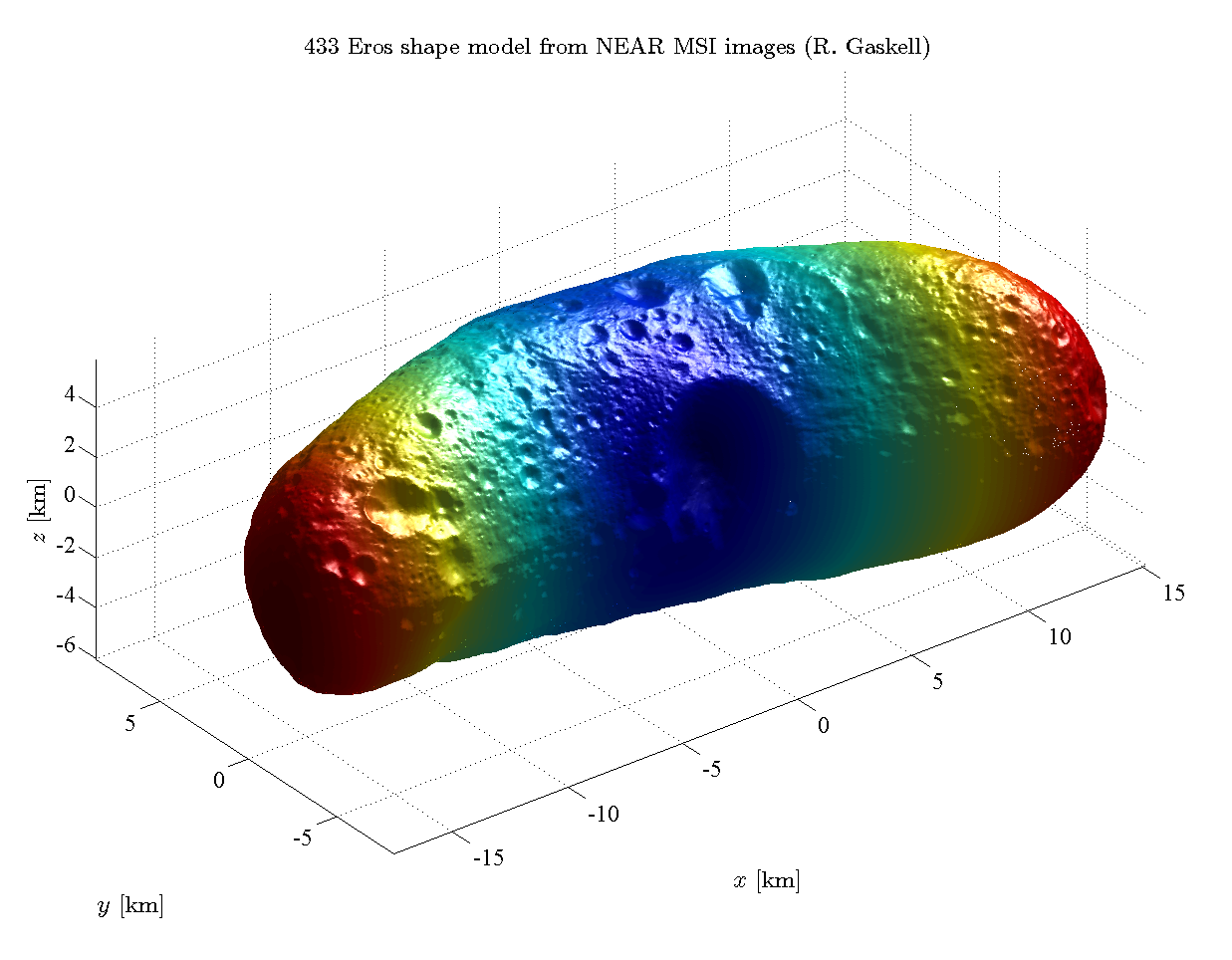}
\caption{Shape model of Eros. Colors from blue to red are representative of increasing distance from Eros barycenter. \href{http://sbn.psi.edu/pds/asteroid/NEAR_A_MSI_5_EROSSHAPE_V1_0/}{Model data: R. Gaskell} \cite{gaskellEros}.\label{fig:Eros433SM}}
\end{center}
\end{figure}

\begin{figure}[ht!]
\begin{center}
\includegraphics[width=0.5\columnwidth]{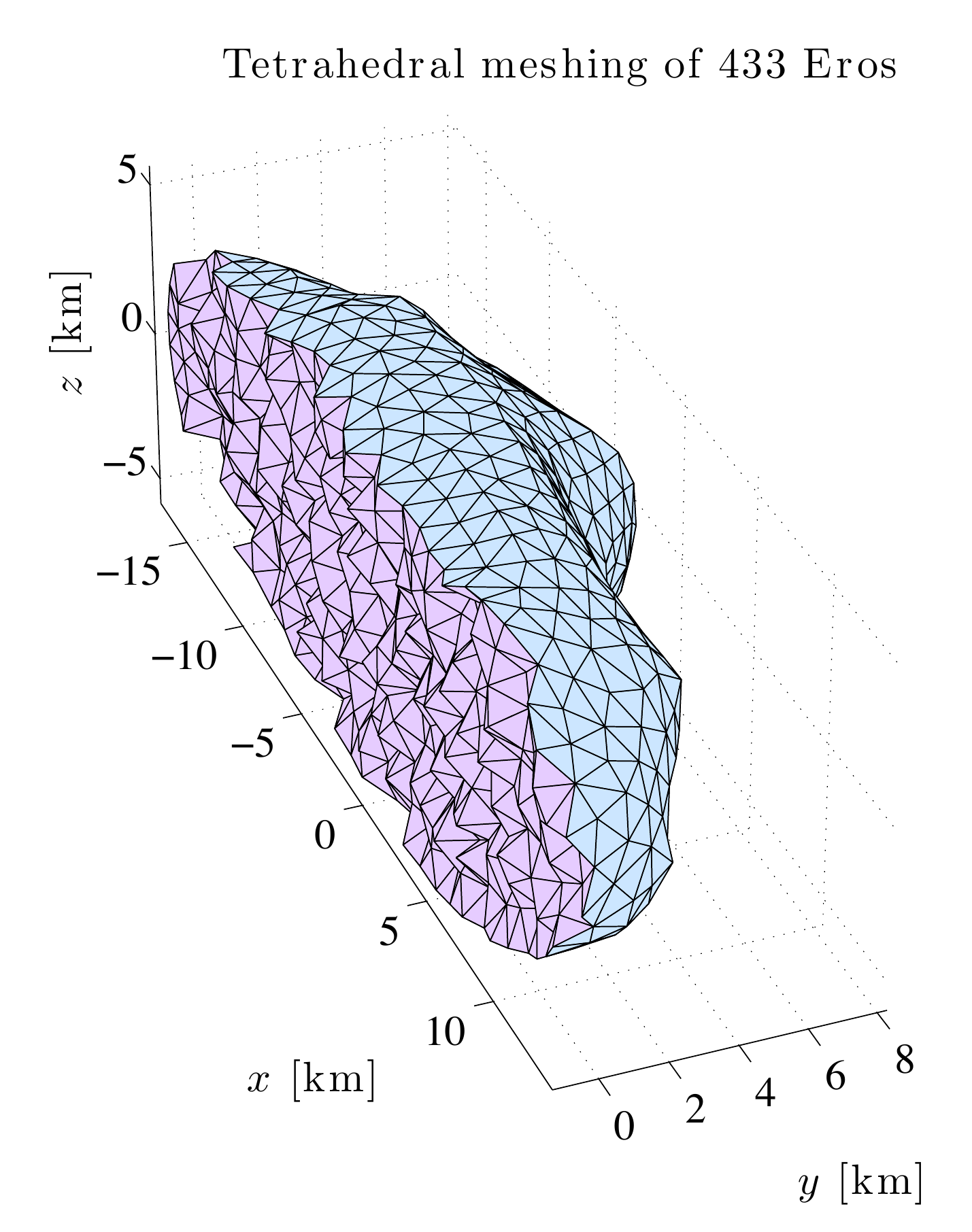}
\caption{Cross-section of tetrahedral meshing of Eros using the modified ``distmesh'' software. Each simplex side length $a \approx 1.2~\mathrm{km}$.}
\label{fig:TetMesh} 
\end{center}
\end{figure}

\begin{figure}[ht!]
\begin{center}
\subfloat[Distribution of tetrahedral side lengths in Eros meshing.
]{\includegraphics[width=0.475\textwidth]{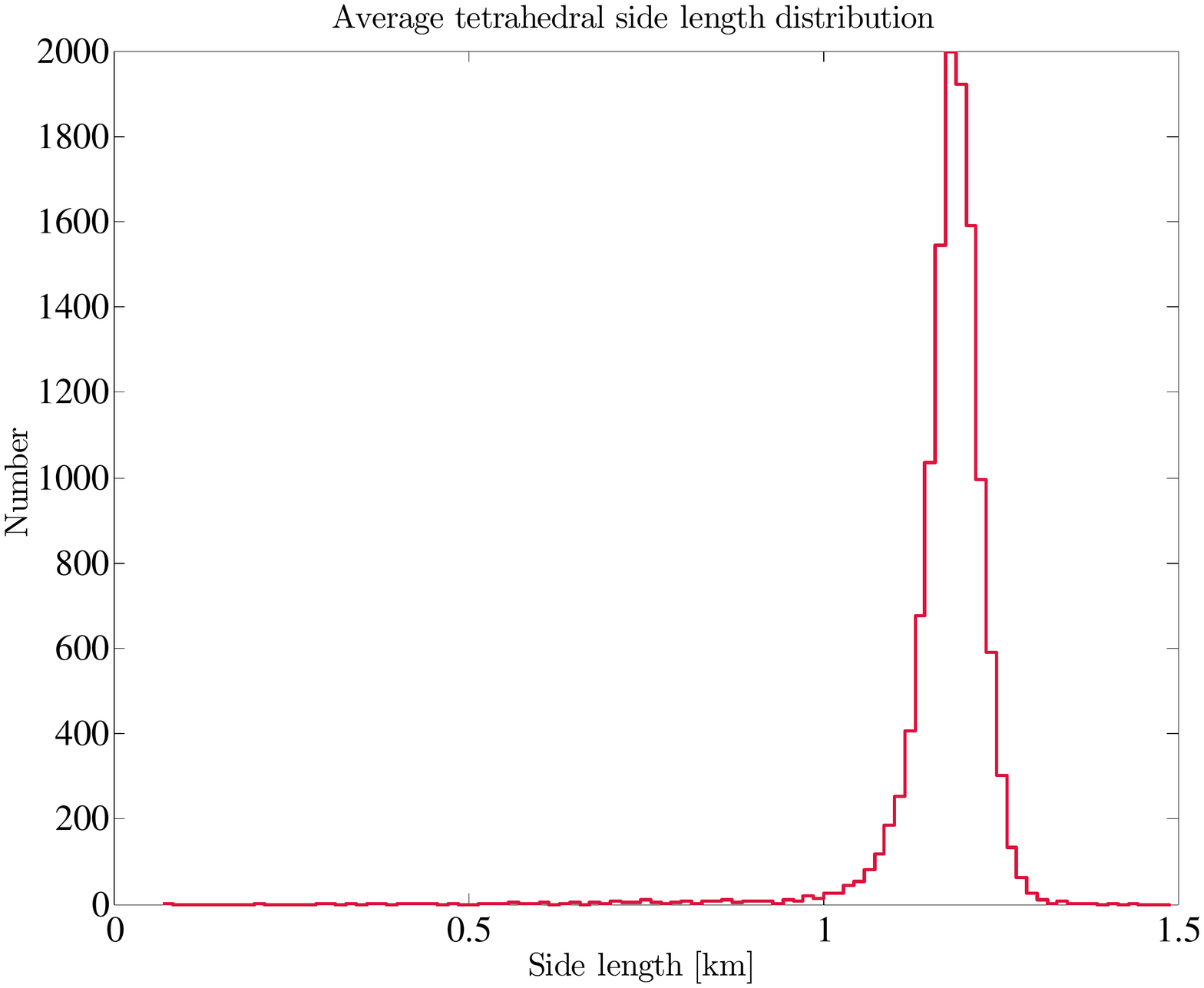} \label{fig:TetSideDist}}
\subfloat[Distribution of tetrahedral volumes in Eros meshing.
]{\includegraphics[width=0.475\textwidth]{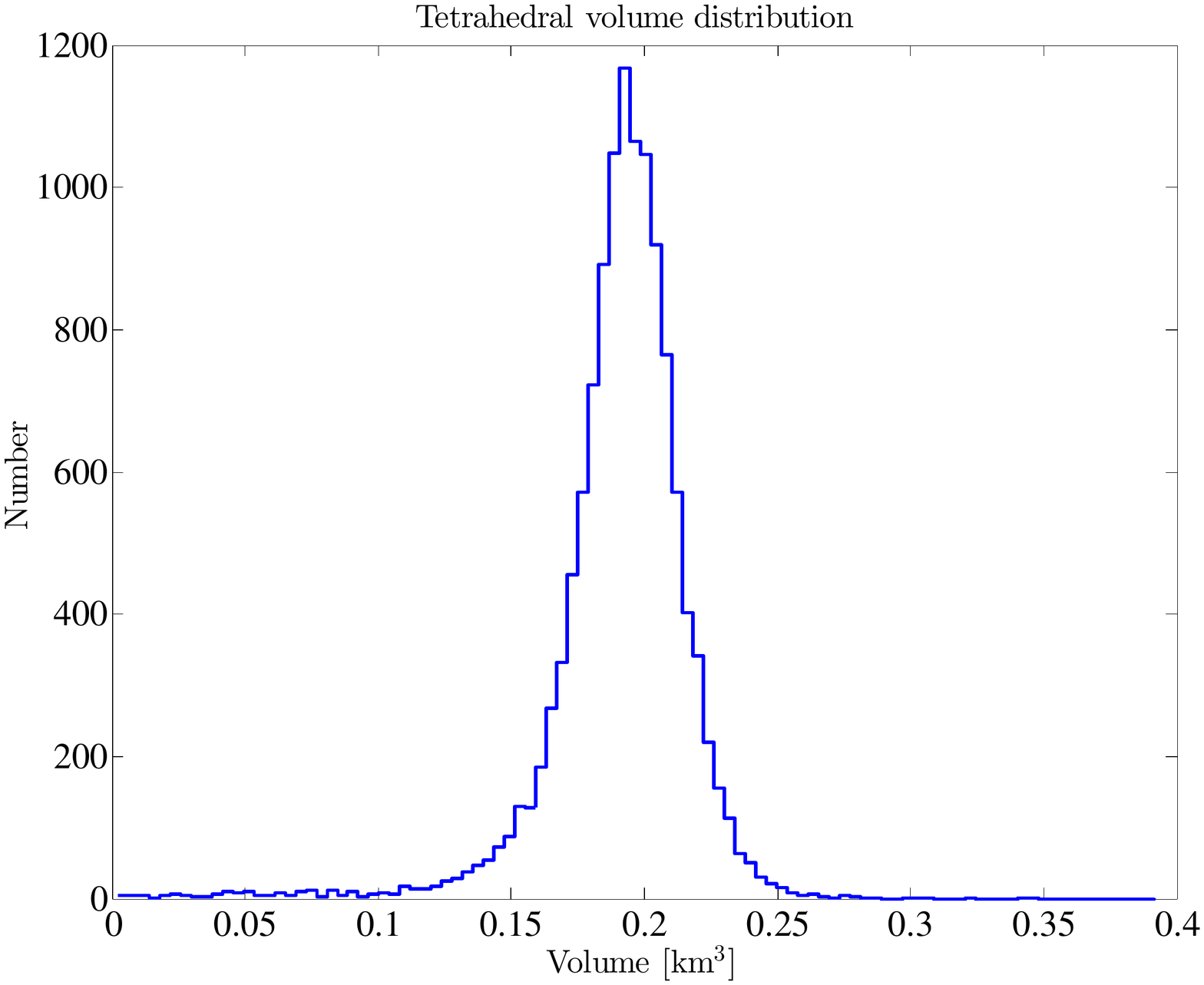} \label{fig:TetVolDist}}
\end{center}
\caption{Characteristics of Eros meshing.}
\label{fig:TetDist}
\end{figure}

\begin{figure}[ht!]
\begin{center}
\includegraphics[width=0.90\textwidth]{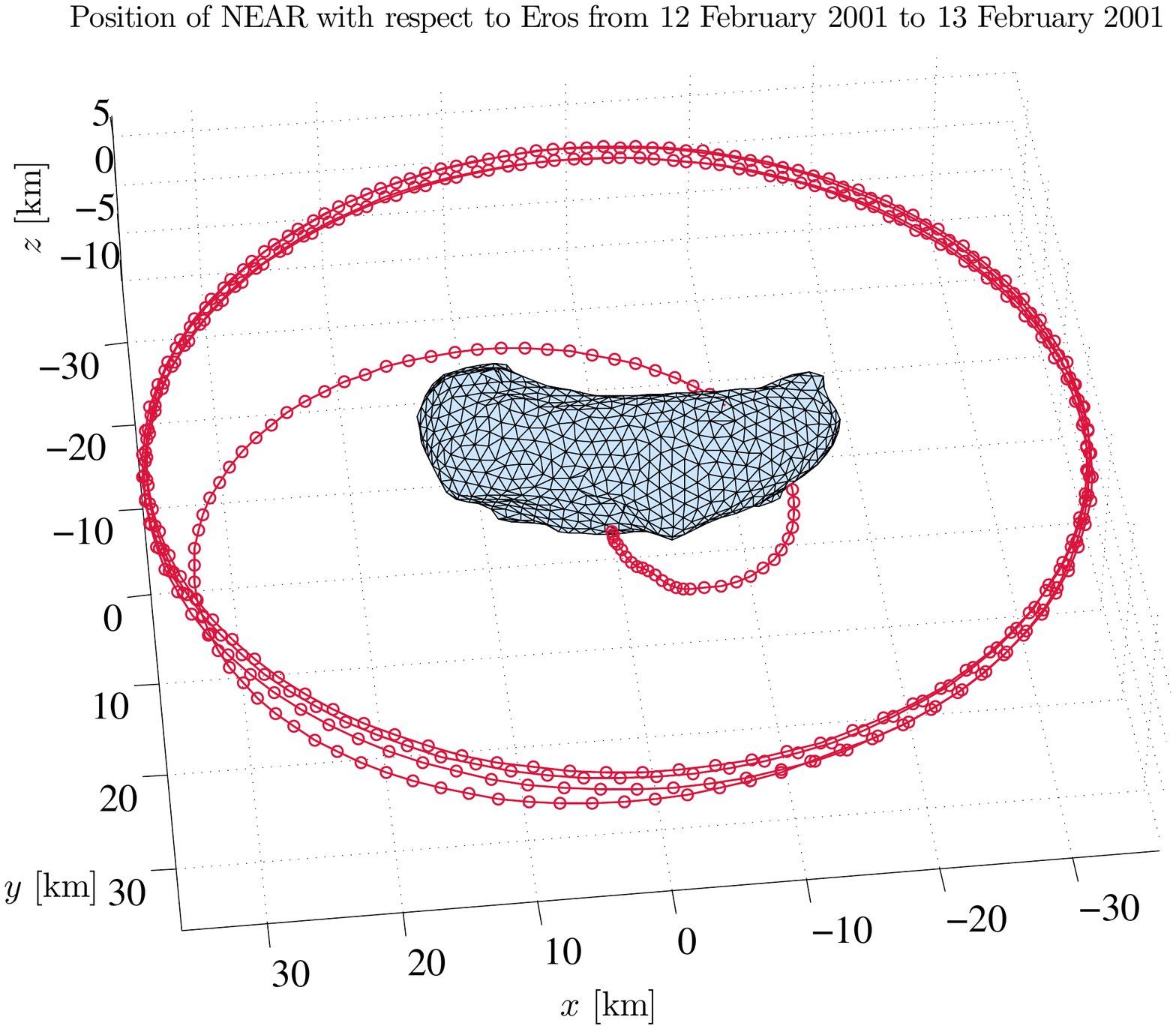}
\caption{Demonstration of the fidelity of our Eros mesh with respect to the NEAR trajectory in the final day of its orbit up to landing.}
\label{fig:Tracking_in} 
\end{center}
\end{figure}

To generate a meaningful simulation of Eros's magnetic field, we use the Eros shape model data in order to develop a volumetric meshing of Eros. The shape model data is from R. Gaskell \cite{gaskellEros} (Fig. \ref{fig:Eros433SM}). To mesh the interior of the asteroid, we use the freely-available ``distmesh'' software \cite{persson2004simple}. The software was modified appropriately to account for the Eros shape geometry using custom scripts. The meshing allows us to populate the interior of the asteroid with roughly uniform tetrahedra (Fig. \ref{fig:TetMesh}). In this implementation, we find that the meshing, while not resulting in all simplectic volumes being the same, nevertheless results in a distribution that clusters around a well-defined mean, with an average side length of about $1.2~\mathrm{km}$, and an average volume of about $0.2~\mathrm{km^3}$ (Fig. \ref{fig:TetDist}). The meshing results in approximately 12,300 volume elements. To demonstrate the fidelity of our meshing procedure, we show in Fig. \ref{fig:Tracking_in} the position of NEAR with respect to Eros during the final day of its orbit lifetime, up until it lands on the surface.

\subsection{Magnetization of Volume Elements}
After meshing the asteroid, we then consider the levels of magnetization which we assign to each volume element. In general, the dipoles that occupy each volume element can have a different direction and a different magnitude. The strength and direction of the dipole depends on the particular case in question, each of which is summarized in Table \ref{tab:MagCases}; these cases define the range of values we consider in our magnetization phase space exploration. 

\begin{table}[ht!]
\caption{Summary of cases to test.}
\begin{tabular}{|c|c|} \hline
\textbf{Direction} 	&  \textbf{Description}\\ \hline\hline
Nonuniform       		&  $\left<\left(d\bvec{m}/dV'\right)/\left|\left(d\bvec{m}/dV'\right)\right|\right> = (0,0,0)$ \\ \hline
Weakly uniform	    &  $\left(d\bvec{m}/dV'\right)/\left|\left(d\bvec{m}/dV'\right)\right| \approx \left(1,0,0\right)$ \\ \hline 
Very uniform 				&  $\left(d\bvec{m}/dV'\right)/\left|\left(d\bvec{m}/dV'\right)\right| \equiv (1,0,0)$ \\ \hline \hline
\textbf{Magnitude} 	&  \textbf{Description} (NRM units [$\mathrm{A\cdot m^2 \cdot kg^{-1}}$])\\ \hline\hline
Weak             		&  $ \mathrm{NRM} \sim 10^{-6}~\textrm{[Eros, per Acu\~{n}a, \textit{et al.} (2002)]}$ \\ \hline
Medium			    		&  $ 10^{-6}\lesssim \mathrm{NRM} \lesssim 10^{-1}~\textrm{[chondrites, per Wasilewski, \textit{et al.}  (2002)]}$  \\ \hline
Strong			 				&  $ \mathrm{NRM} \sim 10^{-1}~\textrm{[Braille and Gaspra, per Wasilewski, \textit{et al.} (2002)]}$ \\ \hline
\end{tabular}
\label{tab:MagCases}
\end{table}

\begin{figure}[H]
\begin{center}
\includegraphics[width=0.85\textwidth]{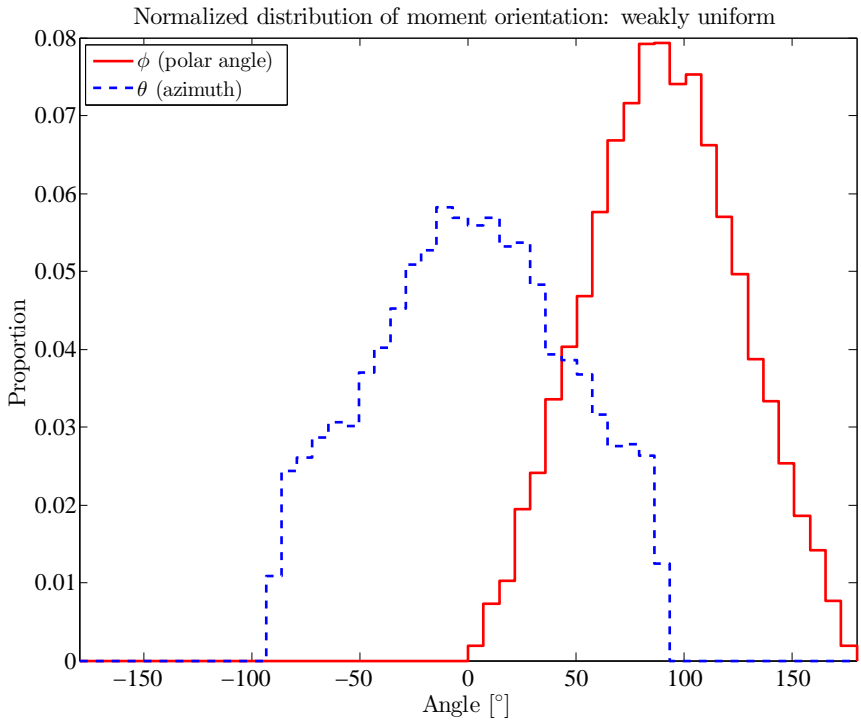}
\caption{Distribution of the dipole orientations $\left(d\bvec{m}/dV'\right)/\left|\left(d\bvec{m}/dV'\right)\right|$ in the weakly uniform case. Orientations are parameterized by polar and azimuthal angles, the polar angle measured with respect to the Eros-fixed $z$-axis (normal to the plane of orbit). In the weakly uniform case, all dipoles are clustered about the positive $x$-axis, with deviations randomly sampled from a normal distribution with $\sigma = 90^{\circ}$. }
\label{fig:DipoleOrient} 
\end{center}
\end{figure}

For a dipole $d\bvec{m}$ occupying a tetrahedron volume element $dV'$, the direction will be given by $\left(d\bvec{m}/dV'\right)/\left|\left(d\bvec{m}/dV'\right)\right|$. In the case in which Eros is composed of totally randomly oriented constituents, we have
\begin{align*}
\left<\frac{\left(d\bvec{m}/dV'\right)}{\left|\left(d\bvec{m}/dV'\right)\right|}\right> &\equiv \int dV' \frac{\left(d\bvec{m}/dV'\right)}{\left|\left(d\bvec{m}/dV'\right)\right|} \\
            &= (0,0,0)
\end{align*}
For the very uniform case, we have all $\left(d\bvec{m}/dV'\right)/\left|\left(d\bvec{m}/dV'\right)\right| = (1,0,0)$, with the $x-y$ plane being that in which the orbit trajectory lies. For the weakly uniform case, we take $\left(d\bvec{m}/dV'\right)/\left|\left(d\bvec{m}/dV'\right)\right| = (1,0,0)$ nominally, but add random variations to the orientation of each dipole. The random variations are chosen from normally-distributed random variables. If we parametrize the dipole direction by a polar and azimuthal angle, then in our simulations, we take the added perturbations to have a mean $0^{\circ}$ and a standard deviation $\sigma = 90^{\circ}$. The direction vectors are suitably normalized. An example of the distribution of weakly-uniform dipoles is shown in Fig. \ref{fig:DipoleOrient}.

In order to assign dipole strengths to each volume element, we consider the range of NRM values given in Wasilewski, \textit{et al.} \cite{wasilewski2002443} and shown in Fig. \ref{fig:MeteoriteSpecimenComps}. Keeping in mind that the values of NRM given for Braille and Gaspra are possibly flawed, we nevertheless use  $\mathrm{NRM} \sim 10^{-1}~\mathrm{A\cdot m^2 \cdot kg^{-1}}$ as an upper limit and $\mathrm{NRM} \sim 10^{-6}~\mathrm{A\cdot m^2 \cdot kg^{-1}}$ as a lower limit. In the strong magnitude case, we take the upper limit value for all volume elements, and in the weak magnitude case, we take the lower limit value for all volume elements; the medium magnitude case is taken as a uniformly random sample of the values between the upper and lower limits. The magnitude of the dipole in a volume element is given by 
\begin{align*}
\left|d\bvec{m}\right| 	&= \left| \frac{d\bvec{m}}{dV'}dV'  \right| \\
												&= \mathrm{NRM} \times \rho dV'.
\end{align*}

\begin{figure}[ht]
\begin{center}
\includegraphics[width=0.9\textwidth]{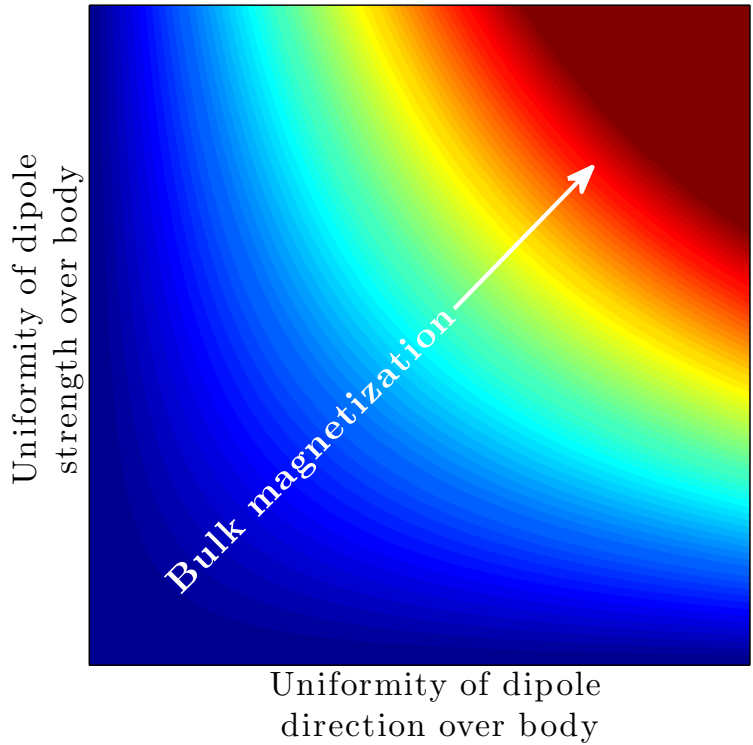}
\caption {Magnetization as a function of dipole strength and direction uniformity over the body.}
\label{fig:Mag_scale} 
\end{center}
\end{figure}

Conceptually, we would expect that the more uniform and the higher intensity the distribution of dipoles within the asteroid, the greater the bulk magnetism we detect (Fig. \ref{fig:Mag_scale}). We ultimately find that this assertion is borne out by our simulation results (Table \ref{tab:MeasNRMLong}).

\subsection{The Role of Orbit in Calculating NRM}\label{sec:OrbitNRM}
We are interested in placing an upper bound on the NRM of the asteroid. Hence, it is essential to consider an orbit that spans only a single plane when assigning an NRM value to the asteroid, in order to remove the possible degeneracy in bulk moment orientation and detected field that would occur if orbits were taken through different planes. By inspection of Eq. \eqref{eq:Hfield}, we see that, by a factor of two, the upper limit of the magnetic field measurement will be achieved when the dipole is oriented with the plane of the orbit. Hence, in order to estimate the $\mathrm{NRM}$ of the asteroid, we take
\begin{align}
\mathrm{NRM}_{\textrm{Eros}}(t) \approx \frac{2 \pi}{\mu_0}\frac{|\bvec{B}(t)|r(t)^3}{M_{\textrm{Eros}}},
\end{align}
where $r$ is the distance to the asteroid (discussed below), and where we have shown explicitly the time dependence of such an evaluation.  

NRM should in principle be independent of the distance through which the measurement is made. As a demonstration into the shape effects intrinsic in making such a calculation, we consider the NRM calculated against both the NEAR orbit to Eros barycenter distance ($r_O$), and against the shortest distance from orbit to a point on Eros's surface ($r_{\mathrm{min}}$). In Fig. \ref{fig:NRM_bounds}, the upper limit (blue line) for $\mathrm{NRM}_{\textrm{Eros}}$ is calculated using $r_{\mathrm{min}}$; the lower bound (broken orange line) is calculated using $r_O$. The value calculated using $r_{\mathrm{min}}$ is consistently higher and reflective of the $r^{-3}$ dependence of the field, the nearer values having a greater influence on the measurements. In the results given below, we use $r_{\mathrm{min}}$ as the baseline against which the bulk NRM is derived and take the overall NRM estimation as

\begin{align}\label{eq:NRM_meandef}
\mathrm{NRM}_{\textrm{Eros}} = \frac{\displaystyle\int_{\textrm{Orbit start}}^{\textrm{Orbit end}} \mathrm{NRM}_{\textrm{Eros}}(t) dt}{t(\textrm{Orbit start}) - t(\textrm{Orbit end})}.
\end{align}

Likewise, for the mean $\left|\bvec{B}\right|$ field measurements simulated, we take an average over the orbit of interest. 

\begin{figure}[ht]
\begin{center}
\subfloat[NEAR $35~\mathrm{km}$ orbit around Eros.
]{\includegraphics[width=0.95\textwidth]{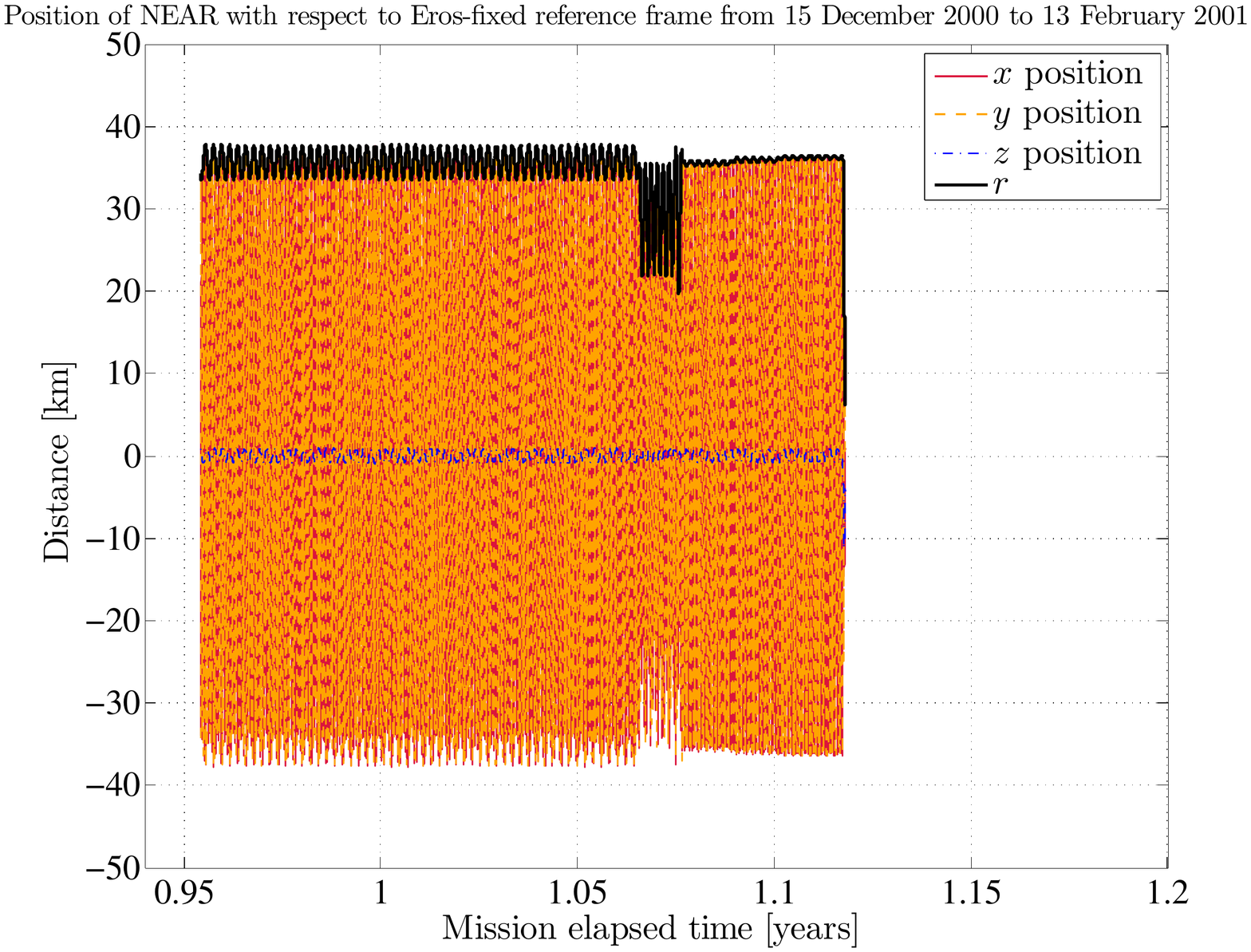} \label{fig:FinalTrajxy}}
\subfloat[Three-dimensional view of orbital period considered.
]{\includegraphics[width=0.95\textwidth]{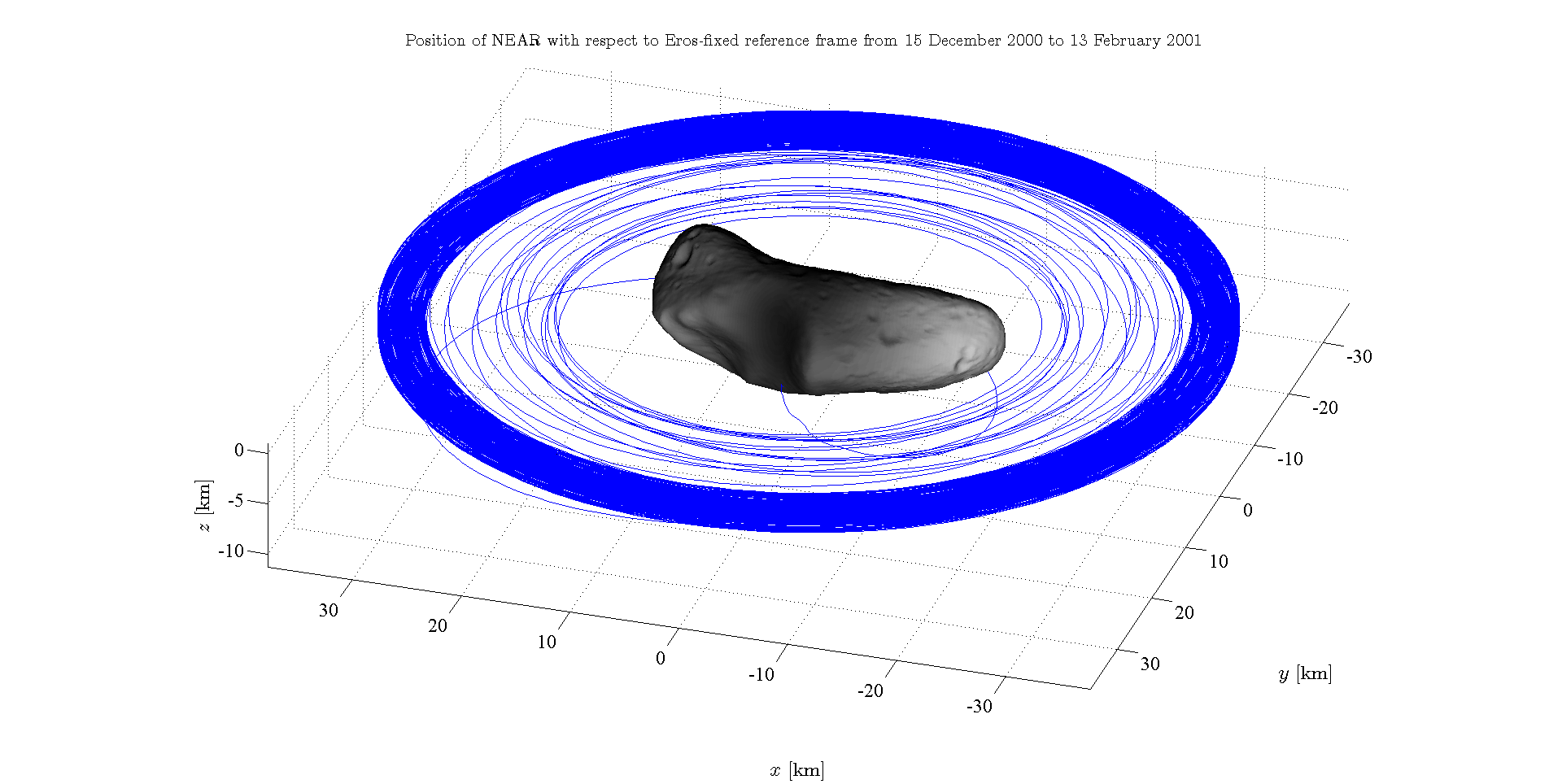} \label{fig:FinalTraj3D}}
\end{center}
\caption{Orbital positions of NEAR in Eros-fixed coordinate system from 15 December 2000 to 13 February 2001. This covers the insertion of the NEAR spacecraft into a $35~\mathrm{km}$ orbit through to its landing on Eros.}
\label{fig:FinalTraj}
\end{figure}

In Fig. \ref{fig:FinalTraj}, we show the orbital case that we have chosen for our simulations. Eros is inserted into a roughly 35-km orbit on 15 December 2001 and lands on Eros on 13 February 2001.

\section{Results}
\subsection{Phase Space Exploration}
In our simulations, for each case study in the phase space exploration, we tabulate both the mean magnitude of the field strength $|\bvec{B}|$ in $\mathrm{nT}$ and the mean upper-bound NRM as given in Eq. \eqref{eq:NRM_meandef}. These results are shown in Tables \ref{tab:MeasNRMLong} and \ref{tab:MeasBfield}. Clearly, as we go through increasing levels of dipole uniformity and dipole strength, we see an increase in bulk magnetism $\mathrm{NRM}_{\textrm{Eros}}$. From a field strength point of view, those of ``medium'' magnitude---randomly sampled from the range $[10^{-6},10^{-1}]~\mathrm{A\cdot m^2 \cdot kg^{-1}}$, and typically having an average NRM of $\sim 2.14 \times 10^{-5}~\mathrm{A\cdot m^2 \cdot kg^{-1}}$---are most compelling for the $\left|\bvec{B}\right|$ values measured by NEAR ($\mathcal{O}\left(1~\mathrm{nT}\right)$). Hence, we use these values as a starting point for the next part of our analysis.

 \begin{table}[ht!]
 \caption{Simulated upper bound NRM values---in [$\mathrm{A\cdot m^2 \cdot kg^{-1}}$]---for Eros. The orbital period ranges from 15 December 2000 to 13 February 2001.}
 \doublespacing{
 \centering
 \begin{tabular}{|c|l|c|c|c|}
 \hline
\multicolumn{2}{|c|}{Simulated} & \multicolumn{3}{c|}{\textbf{Alignment}} \\ \cline{3-5}
\multicolumn{2}{|c|}{NRM values} & \multicolumn{1}{c|}{Nonuniform } & \multicolumn{1}{c|}{Weakly uniform} & \multicolumn{1}{c|}{Very uniform}\\ \hline \hline
 \parbox[t]{4mm}{\multirow{3}{*}{\rotatebox[origin=c]{90}{\textbf{Magnitude}}}} 
 & Weak  			&	$ 6.9 \times 10^{-9}$ &	$ 4.5\times 10^{-7}$			& $ 9.2\times 10^{-7}$			\\ \cline{2-5}
 & Medium 		&	$ 2.4\times 10^{-7}$ 	&	$ 9.3\times 10^{-6}$			& $ 2.0\times 10^{-5}$			\\ \cline{2-5}
 & Strong 		&	$ 8.8\times 10^{-4}$ 	&	$ 4.5\times 10^{-2}$			& $ 9.2\times 10^{-2}$			\\ \cline{2-5}
 \hline
 \end{tabular}
 \label{tab:MeasNRMLong}
 }
 \end{table}

 \begin{table}[ht!]
 \caption{Mean $|\bvec{B}|$---in [$\mathrm{nT}$]---for Eros. The orbital period ranges from 15 December 2000 to 13 February 2001.}
 \doublespacing{
 \centering
 \begin{tabular}{|c|l|c|c|c|}
 \hline
\multicolumn{2}{|c|}{Simulated} & \multicolumn{3}{c|}{\textbf{Alignment}} \\ \cline{3-5}
\multicolumn{2}{|c|}{$|\bvec{B}|$ values} & \multicolumn{1}{c|}{Nonuniform } & \multicolumn{1}{c|}{Weakly uniform} & \multicolumn{1}{c|}{Very uniform}\\ \hline \hline
 \parbox[t]{4mm}{\multirow{3}{*}{\rotatebox[origin=c]{90}{\textbf{Magnitude}}}} 
 & Weak  			&	$0.00019$ &	$0.013$			& $0.026$			\\ \cline{2-5}
 & Medium 		&	$0.0065$ 	&	$0.26$			& $0.56$			\\ \cline{2-5}
 & Strong 		&	$24.99$ 	&	$1261.87$ 	& $2596.96$		\\ \cline{2-5}
 \hline
 \end{tabular}
 \label{tab:MeasBfield}
 }
 \end{table}

\subsection{Deriving Constituent and Bulk NRM values for Eros}
Our goal is to reconstruct possible constituent and bulk Eros NRM values based on the measured field strengths given by Acu\~{n}a, \textit{et al.} The flux density values we consider are $1~\mathrm{nT}$ and $5~\mathrm{nT}$, these values representing the $35~\mathrm{km}$ and surface bounds from the NEAR measurements. Furthermore, since we are interested in finding a realistic case that represents possible bulk magnetization for Eros, we consider the case in which the dipole distribution is weakly uniform directionally. From Table \ref{tab:MeasNRMLong}, we see that the NRM values for the strongly uniform case are typically about a factor of two higher than the weakly uniform case.

\begin{figure}[ht!]
\begin{center}
\includegraphics[width=0.8\textwidth]{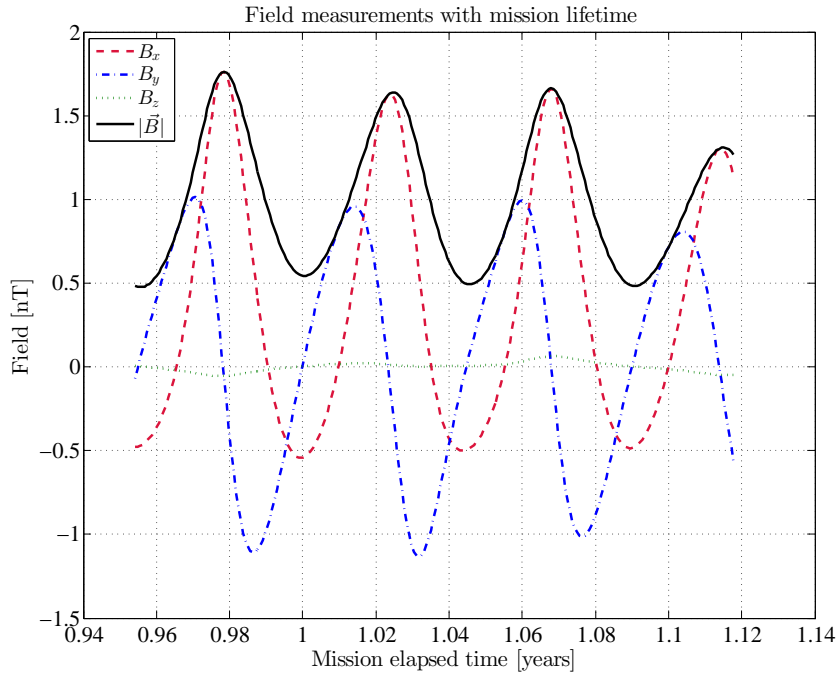}
\caption {Flux density as a function of time for the case in which the mean field strength is $1~\mathrm{nT}$. Constituent NRM is $8\times 10^{-5}~\mathrm{A\cdot m^2 \cdot kg^{-1}}$ and $\mathrm{NRM}_{\textrm{Eros}} = 3.54\times 10^{-5}~\mathrm{A\cdot m^2 \cdot kg^{-1}}$.}
\label{fig:Field} 
\end{center}
\end{figure}

\begin{figure}[ht!]
\begin{center}
\includegraphics[width=1.0\columnwidth]{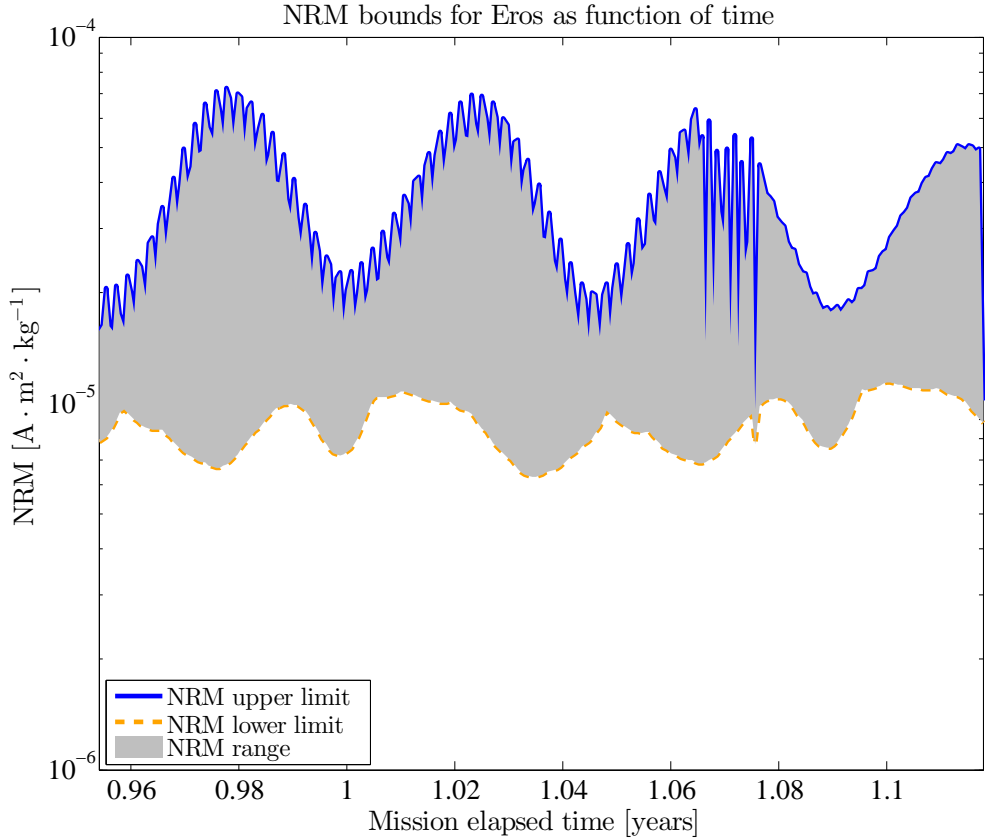}
\caption {NRM range as a function of time for case in which the average field strength is $1~\mathrm{nT}$.}
\label{fig:NRM_bounds} 
\end{center}
\end{figure}

Based on our simulations, we find that a uniform, constituent NRM of $8\times 10^{-5}~\mathrm{A\cdot m^2 \cdot kg^{-1}}$ will lead to an upper limit bulk Erotian NRM of $3.54\times 10^{-5}~\mathrm{A\cdot m^2 \cdot kg^{-1}}$ at a mean flux density of $\approx 1~\mathrm{nT}$. The variation in $\bvec{B}$ with time in this case is shown in Fig. \ref{fig:Field}, and the $\mathrm{NRM}_{\textrm{Eros}}$ values derived from it are shown in Fig. \ref{fig:NRM_bounds}.

For a flux density of $\approx 5~\mathrm{nT}$, we find that a uniformly distributed NRM of $4.0\times 10^{-4}~\mathrm{A\cdot m^2 \cdot kg^{-1}}$ leads to an upper limit Eros bulk NRM of $1.79\times 10^{-4}~\mathrm{A\cdot m^2 \cdot kg^{-1}}$. The results from both cases are summarized in Table \ref{tab:B_lims_NEAR}. 

 \begin{table}[ht!]
 \caption{Summary of constituent and bulk NRM values corresponding to NEAR-measured limiting $\left|\bvec{B}\right|$ field strength values.}
 \doublespacing{
 \centering
 \begin{tabular}{|c|c|c|}
 \hline
Field strength $\left|\bvec{B}\right|$																&	$1.0~\mathrm{nT}$ 		&	$5.06~\mathrm{nT}$   \\ \hline \hline
Constituent NRM $[\mathrm{A\cdot m^2 \cdot kg^{-1}}]$									&	$8\times 10^{-5}$			&	$4.0\times 10^{-4}$  \\	\hline
 $\mathrm{NRM}_{\textrm{Eros}}$ $[\mathrm{A\cdot m^2 \cdot kg^{-1}}]$	&	$3.54\times 10^{-5}$ 	&	$1.79\times 10^{-4}$ \\	\hline
 \end{tabular}
 \label{tab:B_lims_NEAR}
 }
 \end{table}

\section{Discussion}

\begin{figure}
\begin{center}
\includegraphics[width=0.75\columnwidth]{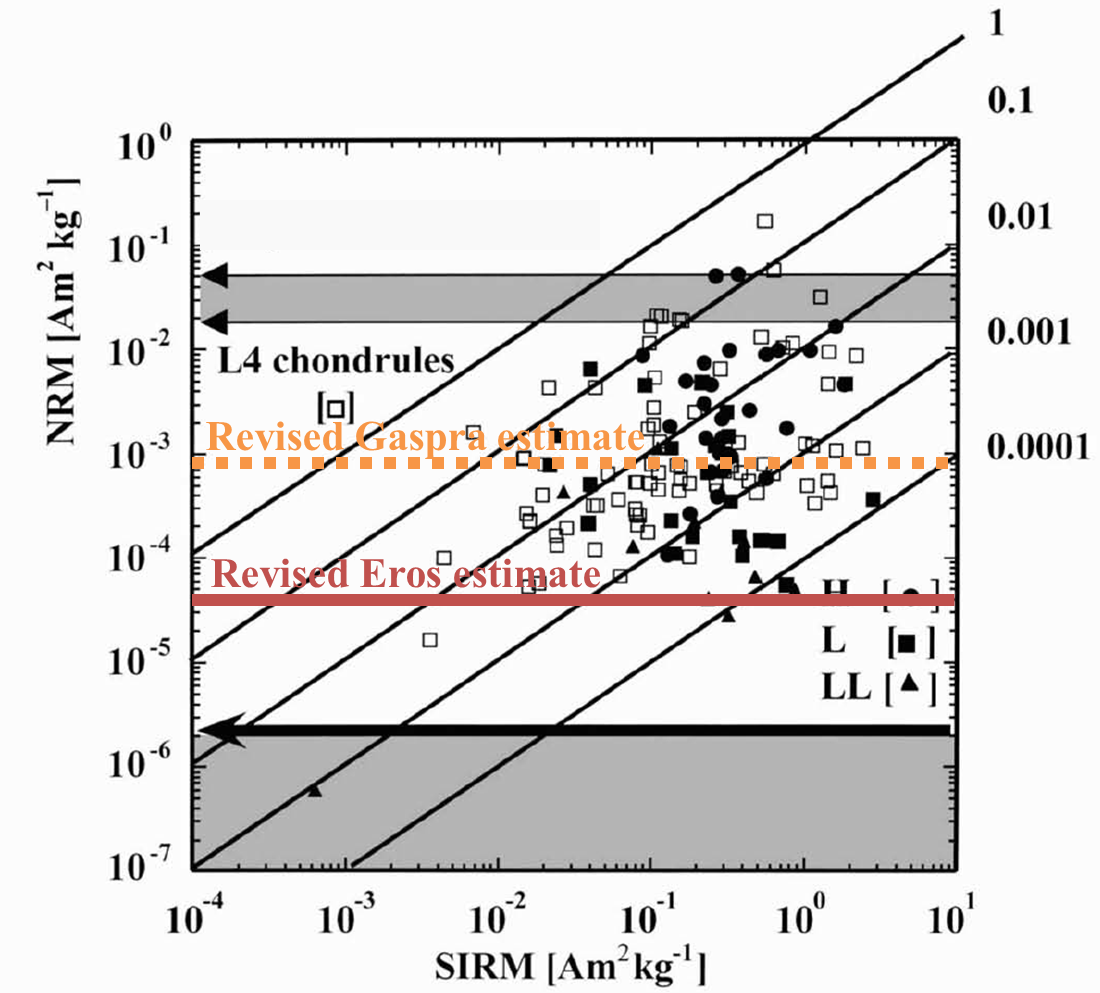}
\caption {Revised bounds on the NRM for both Gaspra and Eros based on this work. Figure adapted from Wasilewski, \textit{et al.} \cite{wasilewski2002443}.}
\label{fig:Rev_Comp} 
\end{center}
\end{figure}

The levels of remanent magnetism that we calculated in the previous section are not inconsistent with that found for other ordinary chondrites, in particular L- and LL-types. The bound that we have placed in the $1~\mathrm{nT}$ case---$3.54\times 10^{-5}~\mathrm{A\cdot m^2 \cdot kg^{-1}}$---is more than an order of magnitude greater than that given by Acu\~{n}a, \textit{et al.}, and the constituent NRM of $8\times 10^{-5}~\mathrm{A\cdot m^2 \cdot kg^{-1}}$ is within the bounds for the specimens given in Wasilewski, \textit{et al.} In Fig. \ref{fig:Rev_Comp}, we summarize our results with the new bounds for Gaspra (discussed above) and Eros as calculated here shown. The fact that we were able to reconstruct a scenario in which Eros was magnetized with constituents that are roughly uniformly oriented, and with dipole moment magnitudes within the range reported for ordinary chondrites suggests that we cannot rule out Eros being magnetized on a global scale. 

With regards to the difference between the bounds we place here and those placed by Acu\~{n}a, \textit{et al.},  Acu\~{n}a, \textit{et al.} derived their upper limit for magnetization by using the bounding field value at Eros's surface ($\sim 5~\mathrm{nT}$) and populating an axisymmetric body with varying values of uniform magnetization throughout, until a value for the magnetization was found ($5\times 10^{-3}~\mathrm{A\cdot m^{-1}}$) that matched a surface flux density of $\sim 5~\mathrm{nT}$. Then, by using the known average density, NRM was determined. We believe, however, that placing a bound requires a more nuanced approach that takes greater account of the influence of orbit and shape, as we have done here. 

\section{Conclusion and Future Work}
We have, by constructing a magnetization model of 433 Eros and taking into account NEAR's orbit, rederived upper bounds on the bulk NRM for Eros. The values that we inferred for the bulk magnetism of Eros were roughly an order of magnitude higher than those reported by Acu\~{n}a, \textit{et al.}, and the levels of magnetism for its constituents were within the bounds of NRM values reported for L- and LL-type chondrites. We therefore tentatively suggest that global magnetization of Eros is possible, and note that we cannot rule out a scenario in which Eros originated from, as one picture of asteroid and meteorite origin suggests, crustal aggregate on a differentiated parent body. 

There are numerous opportunities to extend this work to more complex physical pictures. In particular, we may use the shape model and magnetization-generation modules and incorporate solar-wind/IMF interactions to recreate a Gaspra-type situation. We note again that one of the challenges associated with attempting to assess the magnetization of a particular sample are the scales on which magnetization can be present. In our models, we have discretized the asteroid into tetrahedra with average side length of $1.2~\mathrm{km}$. In future implementations, it may be worthwhile to consider finer-scale effects, particularly during the landing portion of the mission. There are certainly computational difficulties intrinsic to this, but by focusing on specialized regions, it may be possible to glean more information about the influence of localized variation in magnetism for Eros. 

\section*{Acknowledgements}
The author would like to extend thanks to Anton Ermakov for pointing out that he missed a row in the original Gaskell shape model data set. The author would also like to thank Ben Weiss for his comments on the draft of this paper and initial recommendations. 

\newpage
\bibliographystyle{unsrt}
\bibliography{Eros_References}

\newpage
\appendix
\section{Code Organization}
\begin{figure}[ht]
\begin{center}
\includegraphics[width=0.75\textwidth]{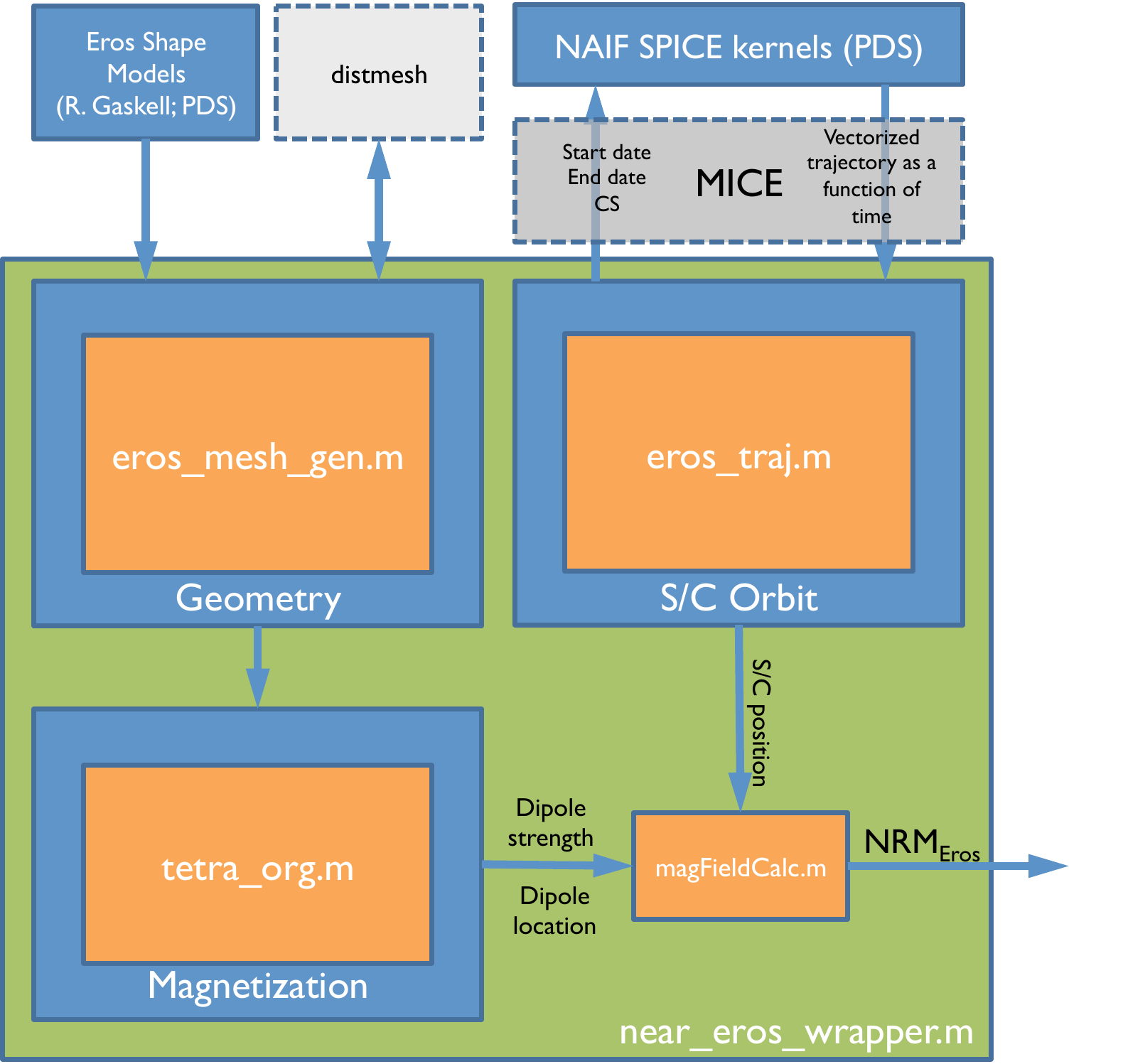}
\caption {Organization of simulation structure.}
\label{fig:Code_org} 
\end{center}
\end{figure}

All simulations for this study were carried out using MATLAB software \cite{MATLAB2012}. SPICE kernel data were read in using the MATLAB implementation of CSPICE, which is freely available through the NASA Navigation and Ancillary Information Facility (NAIF). The structure of the code is shown in Fig. \ref{fig:Code_org}.

\end{document}